\documentclass[3p,times]{elsarticle}

\usepackage{ecrc}

\usepackage{epstopdf}

\volume{00}

\firstpage{1}

\journalname{Nuclear Physics A}

\runauth{Sanghoon Lim}

\jid{nupha}

\jnltitlelogo{Nuclear Physics A}

\usepackage{graphicx}
\usepackage{amsmath,amssymb}

\usepackage{xspace}
\newcommand{\dau}{\mbox{$d$$+$Au}\xspace}
\newcommand{\pt}{\mbox{$p_T$}\xspace}
\newcommand{\kt}{\mbox{$k_T$}\xspace}
\newcommand{\auau}{\mbox{Au$+$Au}\xspace}
\newcommand{\cucu}{\mbox{Cu$+$Cu}\xspace}
\newcommand{\rda}{\mbox{$R_{dA}$}\xspace}
\newcommand{\raa}{\mbox{$R_{AA}$}\xspace}
\newcommand{\rcp}{\mbox{$R_{CP}$}\xspace}
\newcommand{\ncoll}{\mbox{$\langle N_{\rm coll}\rangle$}\xspace}
\newcommand{\npart}{\mbox{$\langle N_{\rm part}\rangle$}\xspace}

\newcommand{\sqsntwo}{\mbox{$\sqrt{s_{_{NN}}}=200$~GeV}\xspace}

\newcommand{\pp}{\mbox{$p$$+$$p$}\xspace}

\begin{document}

\begin{frontmatter}



\title{Measurement of electrons from heavy-flavor decays from \pp, \dau, and \cucu~collisions in the PHENIX experiment}

\author{Sanghoon Lim (for the PHENIX\fnref{col1} Collaboration)}
\fntext[col1] {A list of members of the PHENIX Collaboration and acknowledgements can be found at the end of this issue.}
\address{Physics Department, Yonsei University, Seoul 120-749, Republic of Korea}




\begin{abstract}
Charm and bottom quarks are formed predominantly by gluon fusion in the initial hard scatterings at RHIC, making them good probes of the full medium evolution.
Previous measurements at RHIC have shown large suppression and azimuthal anisotropy of open heavy-flavor hadrons in \auau collisions at \sqsntwo.
Explaining the simultaneously large suppression and flow of heavy quarks has been challenging. 
To further understand the heavy-flavor transport in the hot and dense medium, it is imperative to also measure cold nuclear matter effects which affect the initial distribution of heavy quarks as well as the system size dependence of the final state suppression. 
In this talk, new measurements by the PHENIX collaboration of electrons from heavy-flavor decays in \pp, \dau, and \cucu collisions at \sqsntwo are presented.
In particular, a surprising enhancement of intermediate transverse momentum heavy-flavor decay leptons in \dau at mid and backward rapidity are also seen in mid-central \cucu collisions.
This enhancement is much larger than the expectation from anti-shadowing of the parton distributions and is theoretically unexplained.
\end{abstract}


\end{frontmatter}



\section{Introduction}
\label{intro}
Heavy quarks, charm and bottom, are produced in the initial hard scattering. Therefore, they are good probes to study the evolution of the medium produced in heavy ion collisions.
In addition, the predominant process of heavy quark production at RHIC energy, gluon fusion, is sensitive to initial state modifications such as shadowing.
The PHENIX experiment has excellent capabilities to measure leptons from heavy-flavor decay both in the central arms at mid-rapidity ($|\eta|<0.35$) and the muon arms at forward rapidity ($1.2<|\eta|<2.2$). 
The previous PHENIX results in \pp collisions, considered as a baseline measurement for the results from other collisions, show a good agreement with FONLL calculation within uncertainties~\cite{HFe_pp}.
In order to quantify medium effects compared to the \pp collisions, the nuclear modification factor is defined as the ratio of invariant yields scaled by the number of binary collisions for a certain centrality range (\ncoll), 
\begin{equation}
R_{AA~(dA)} = \frac{1}{\ncoll} \cdot \frac{dN_{AA~(dA)}/dp_{T}}{dN_{pp}/dp_{T}}.
\end{equation}
In central \auau collisions, where the number of binary collisions is almost 1000, a large suppression of heavy-flavor decay electron production compared to the scaled \pp result is observed~\cite{HFe_auau}.
Even though it is clear that suppressing effects are dominant in the largest collision system at RHIC, it is still important to understand other nuclear effects which are also convoluted in.
Since RHIC has the flexibility to use various beam species, we can study which nuclear effects are dominant in a certain size of collision system such as \dau and \cucu.

\section{\dau collisions}
\begin{figure}
\begin{center}
\includegraphics*[width=0.45\textwidth]{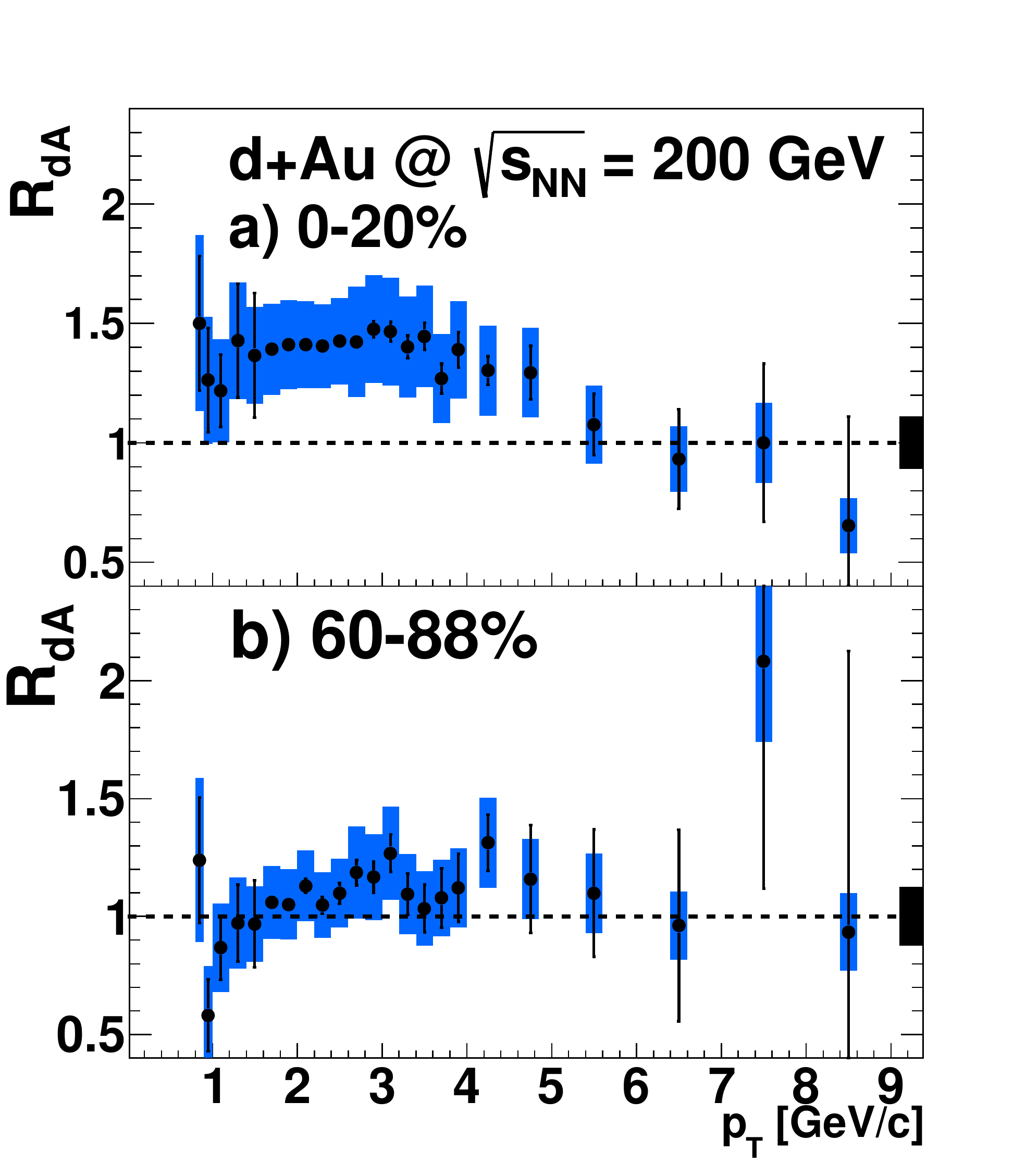}\\
\caption{\rda as a function of \pt of heavy-flavor electrons at mid-rapidity in the most central (top) and the most peripheral (bottom) centrality classes.}
\label{fig:HFe_dau}
\end{center}
\end{figure}

The \dau collision system is considered as a control experiment to study initial state effects, because suppressing effects from the hot and dense medium in heavy ion collisions can be minimized in this low multiplicity collision system.
PHENIX has measured heavy-flavor decay electrons at mid-rapidity~\cite{HFe_dau} and heavy-flavor decay muons at forward ($d$-going direction) and backward (Au-going direction) rapidity~\cite{HFmu_dau} based on large statistics of data collected in 2008.
Figure~\ref{fig:HFe_dau} shows \rda as a function of \pt for heavy-flavor decay electrons in the most central (top) and most peripheral (bottom) centrality classes measured at mid-rapidity~\cite{HFe_dau}.
Heavy-flavor decay electron production is clearly enhanced at moderate \pt region in central \dau collisions ($\ncoll\approx15.1$), however almost no modification ($\rda\sim1$) is observed in peripheral \dau collisions ($\ncoll\approx3.2$).
By comparing with the central \auau results, we can determine that suppressing effects are dominant only in heavy ion collisions.

It is interesting to compare the \dau results from different rapidities, because gluons of different momentum fractions in the Au nucleus, such as those within the shadowing and the anti-shadowing regions, can be accessed in different rapidity regions. 
The heavy-flavor decay muon measurement at backward rapidity shows a similar enhancement with the results at mid-rapidity, but a suppression is seen only at forward rapidity~\cite{HFmu_dau}.
One interesting observation is that any model calculation considering initial state effects, i.e, modification of nuclear parton distribution function and initial \kt broadening, can not simultaneously reproduce the data at forward and backward rapidity.
Furthermore, there is no current theoretical approach with initial state effects to successfully explain the enhancement of heavy-flavor decay electron production seen at mid-rapidity.
Recently, PHENIX has measured angular anisotropies and long-range correlations in central \dau collisions at RHIC energy~\cite{dau_flow1, dau_flow2}, and a model calculation considering radial flow qualitatively reproduces the enhancement of heavy-flavor decay electron production~\cite{HFe_dau_flow}.
From these facts, the enhancement of heavy-flavor decay lepton production seen at mid and backward rapidity in central \dau collisions could possibly result from final state interaction. 

\section{\cucu collisions}
\begin{figure}
\begin{center}
\includegraphics*[width=0.30\textwidth]{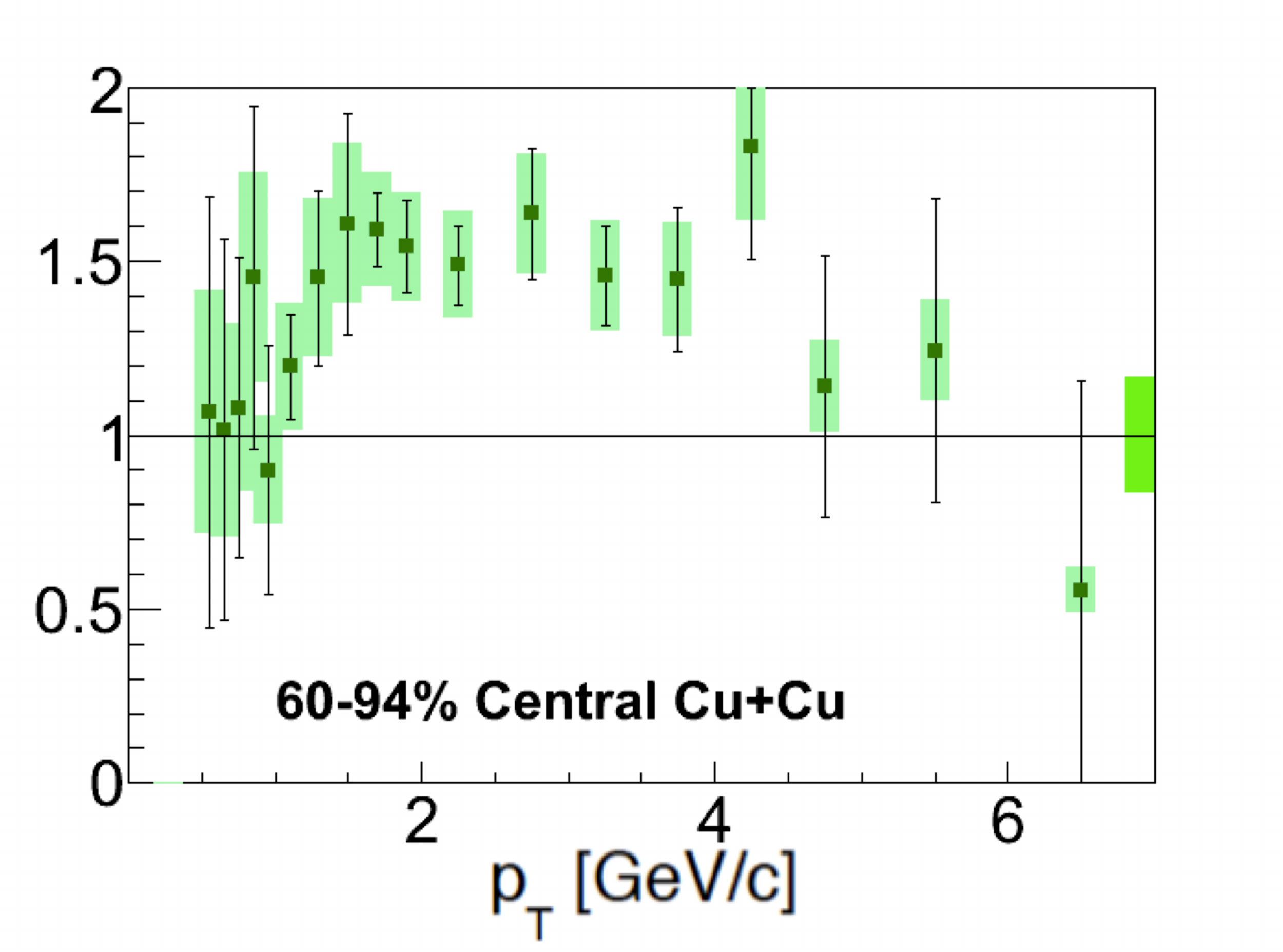}
\includegraphics*[width=0.30\textwidth]{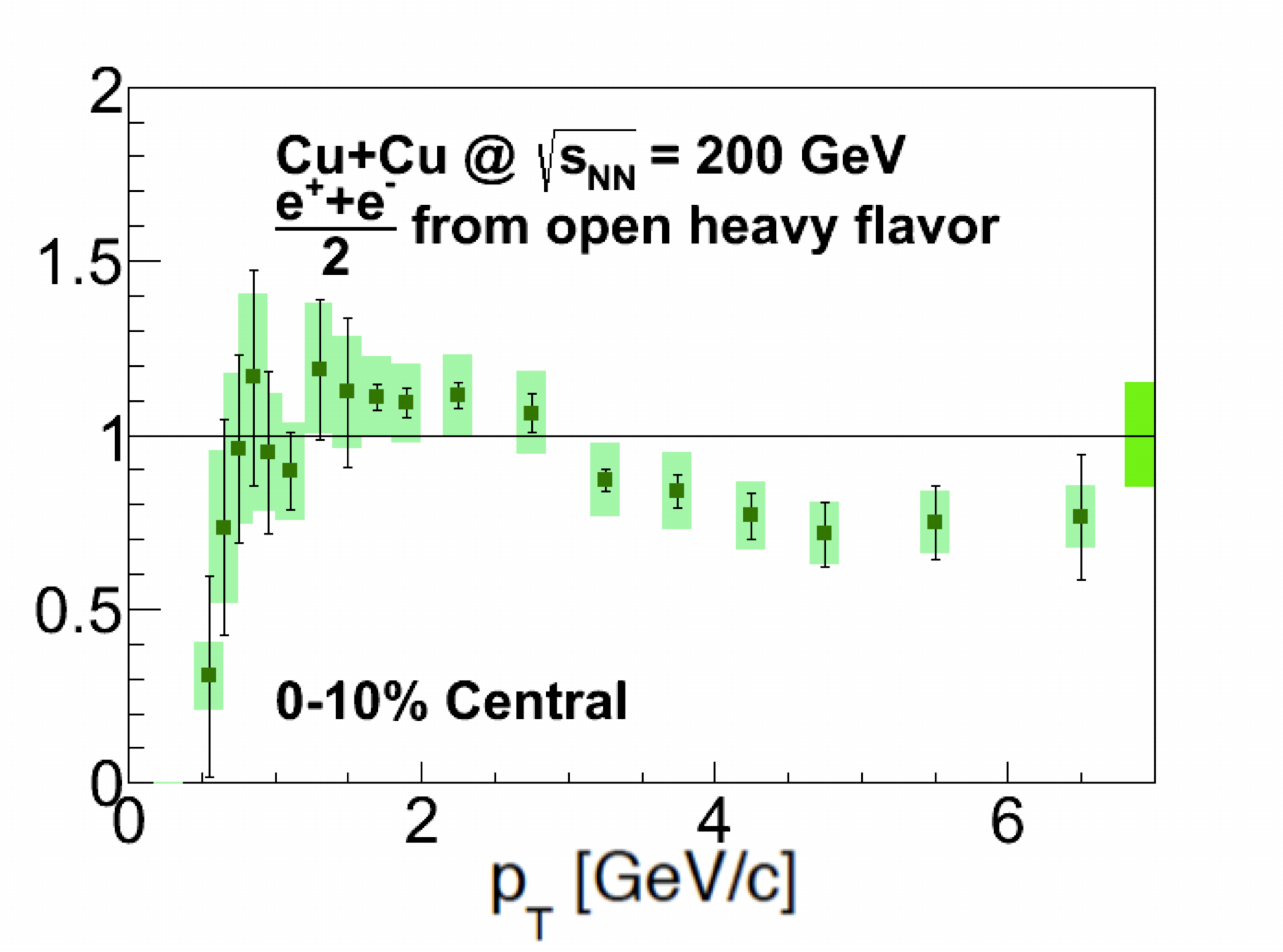}
\includegraphics*[width=0.34\textwidth]{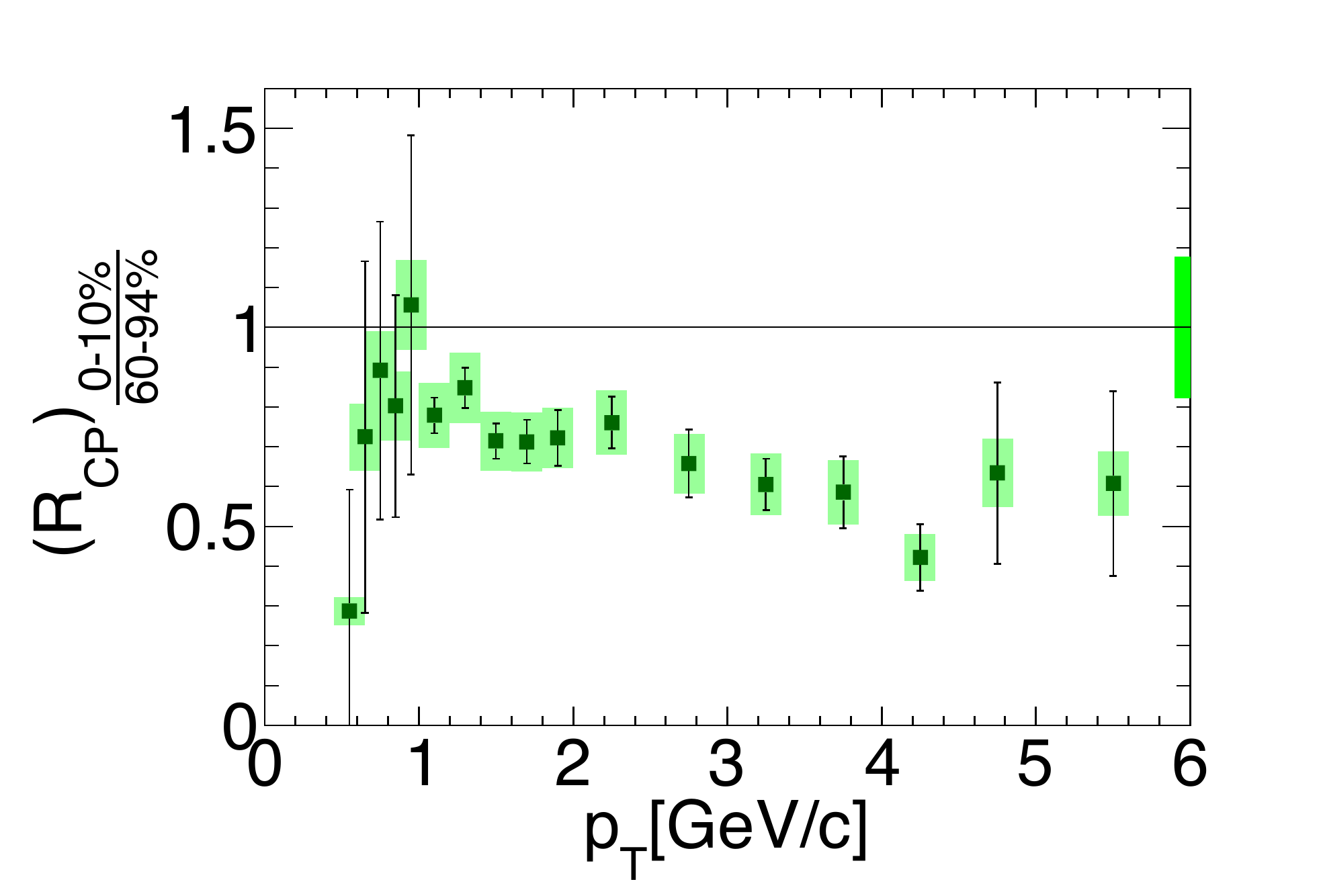}\\
\caption{\raa of heavy-flavor decay electrons in peripheral (left) and central (center) \cucu collisions and \rcp (right)}
\label{fig:HFe_cucu}
\end{center}
\end{figure}

The \cucu collision system size is in between the \dau and \auau collision systems, and heavy-flavor decay electron measurements in this intermediate system can provide a connection between the enhancement in \dau collisions and the suppression in \auau collisions.
Based on large statistics data collected in 2005, PHENIX has measured heavy-flavor decay electrons in various centrality classes at mid-rapidity~\cite{HFe_cucu}.
Figure~\ref{fig:HFe_cucu} shows \raa in the most peripheral (left) and central (center) centrality classes and \rcp (right) of heavy-flavor electrons in \cucu collisions where \rcp is defined as,
\begin{equation}
R_{CP} = \frac{N_{coll}^{peripheral}}{N_{coll}^{cent}} \cdot \frac{dN_{AA}^{central}/dp_{T}}{dN_{AA}^{peripheral}/dp_{T}}.
\end{equation}
In peripheral \cucu collisions ($\ncoll\approx5.1$), heavy-flavor decay electron production is clearly enhanced in the intermediate \pt region, which is similar to the results in central \dau collisions, and a slight suppression is observed at $\pt<3~{\rm GeV}/c$ in central \cucu collisions ($\ncoll\approx182.7$).
A comparison between the most central and peripheral results (\rcp), which takes into account the enhancement in peripheral \cucu collisions, shows a clear suppression of heavy-flavor decay electron production.
Therefore, we can conclude that suppressing effects are dominating in central \cucu collisions, whereas there is a clear enhancement in peripheral \cucu collisions. 

\section{System size dependence}

\begin{figure}
\begin{center}
\includegraphics*[width=0.45\textwidth]{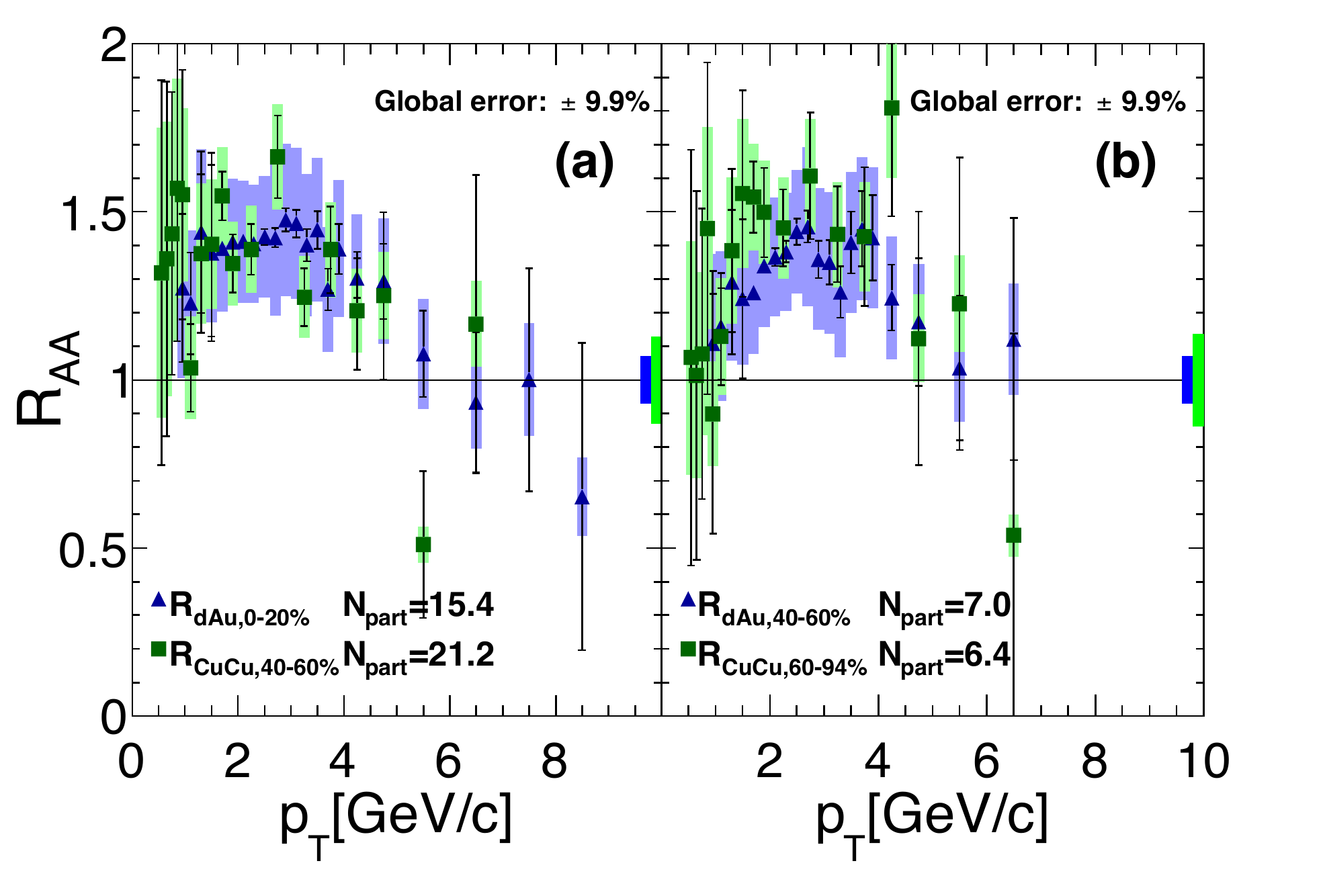}
\includegraphics*[width=0.45\textwidth]{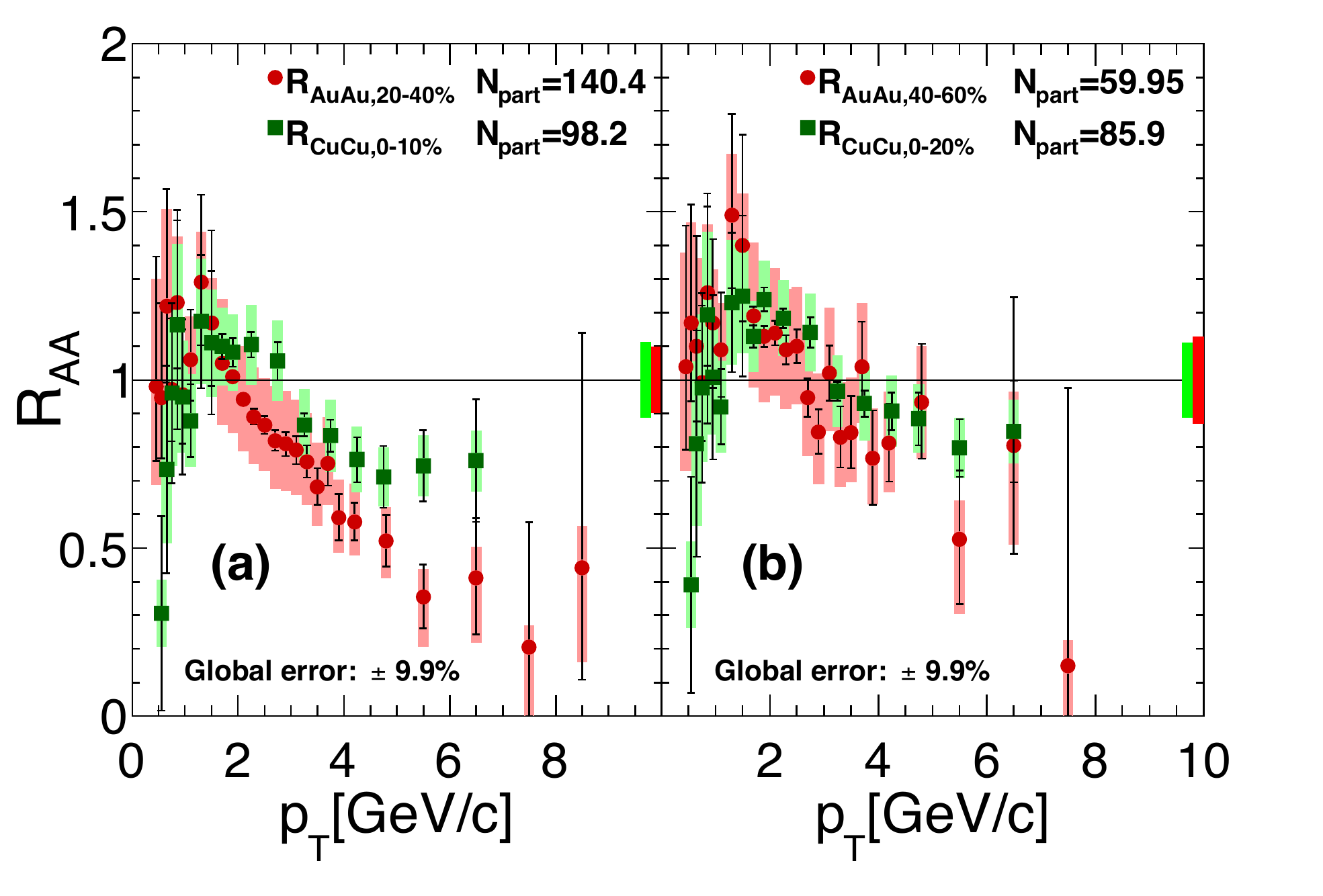}\\
\caption{Comparison of \raa between \dau and \cucu (left) and \cucu and \auau (right) of similar system sizes (\npart).}
\label{fig:HFe_comp}
\end{center}
\end{figure}

Based on the heavy-flavor decay electron measurements in various collision systems, we can see how modifications (nuclear effects) evolve as system size grows. 
Figure~\ref{fig:HFe_comp} shows a comparison of \raa as a function of \pt between selections of \dau, \cucu, and \auau systems with similar number of participants (\npart).
The two left panels show comparisons of \raa between central \dau and peripheral \cucu collisions.
Both \dau and \cucu results show a similar trend of enhancement.
In the two right panels, \raa from central \cucu and mid-central \auau collisions also show similar \pt shapes of heavy-flavor decay electron production compared to the same \pp reference.

\begin{figure}
\begin{center}
\includegraphics*[width=0.8\textwidth]{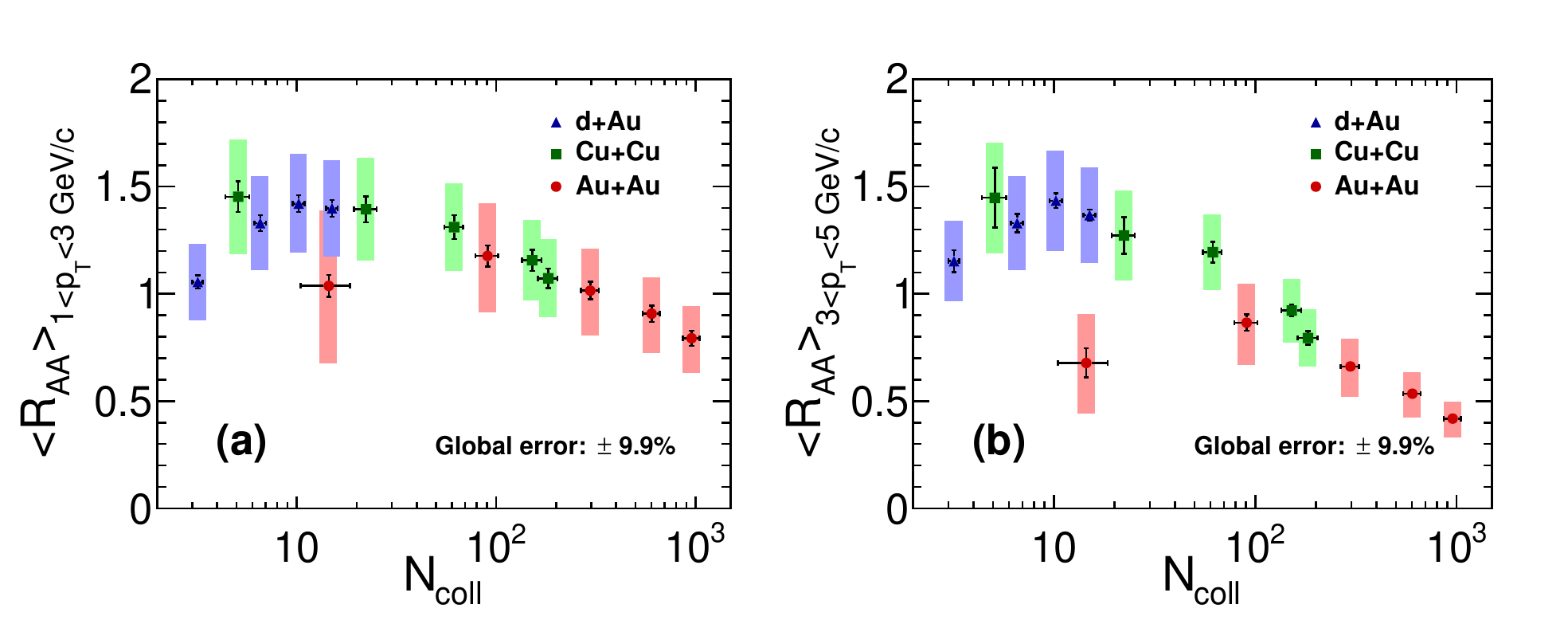}\\
\caption{\raa of heavy-flavor electrons as a function of \ncoll in two \pt ranges for various collision systems.}
\label{fig:HFe_Npart}
\end{center}
\end{figure}

In order to look at a trend of the entire collision system, Figure~\ref{fig:HFe_Npart} shows \raa as a function of \ncoll measured at mid-rapidity for two \pt ranges in \dau, \cucu, and \auau collision systems.
A reasonably smooth trend is seen from \dau and peripheral \cucu collisions, where enhancement effects are dominating, to central \cucu and \auau collisions, where suppression effects take over.

\section{Summary}
PHENIX has measured heavy-flavor decay leptons in various collision systems, \pp, \dau, \cucu, and \auau collisions, across a wide rapidity range.
The heavy-flavor decay electron results at mid-rapidity show a trend of decreasing \raa with increasing system size from the enhancement seen in \dau and peripheral \cucu collisions to suppression in \auau collisions.
The enhancement seen in central \dau collisions at mid and backward rapidity regions is larger than the expectation from initial state effects, and a model calculation inspired by recent results of hydrodynamic behavior in \dau collisions raises the possibility of final state interaction.
A new silicon vertex tracking system (VTX and FVTX) has measured a very large \auau dataset in 2014.
Based on precise vertex position information, new measurements such as $D/B$ separation are expected, and these results will help to understand the production and modification of charm and bottom quarks.








\end{document}